\documentclass[12pt]{iopart}
\usepackage{graphicx}
\usepackage{epsfig}
\usepackage{amsfonts}
\usepackage{amssymb}

\begin{document} 

\title{Punchets: nonlinear transport in Hamiltonian pump-ratchet hybrids} 

\author{Thomas Dittrich, Nicol\'as Medina S\'anchez}

\address{Departamento de F\'\i sica, Universidad Nacional de Colombia, Bogot\'a~D.C., Colombia}
\ead{tdittrich@unal.edu.co}


\begin{abstract}
``Punchets'' are hybrids between ratchets and pumps, combining a spatially periodic static potential,
typically asymmetric under space inversion, with a local driving that breaks time-reversal invariance, and are intended to model metal or semiconductor surfaces irradiated by a collimated laser beam. 
Their crucial feature is irregular driven scattering between asymptotic regions supporting periodic
(as opposed to free) motion. With all binary spatio-temporal symmetries broken, scattering in
punchets typically generates directed currents. We here study the underlying nonlinear transport
mechanisms, from chaotic scattering to the parameter dependence of the currents, in three types
of Hamiltonian models, (i) with spatially periodic potentials where only in the driven scattering
region, spatial and temporal symmetries are broken, and (ii), spatially asymmetric (ratchet)
potentials with a driving that only breaks time-reversal invariance. As more realistic models
of laser-irradiated surfaces, we consider (iii), a driving in the form of a running wave confined 
to a compact region by  a static envelope. In this case, the induced current can even run against
the direction of wave propagation, drastically evidencing of its nonlinear nature. Quantizing punchets
is indicated as a viable research perspective.
\end{abstract}

\maketitle

\section{Introduction}\label{intro}

Directed transport induced by nonlinear dynamics has been studied in two alternative settings,
in periodically driven sawtooth potentials (``ratchets'') \cite{Rei02} and in the framework of driven
chaotic scattering (``pumps'') \cite{AG99}. The concept of ratchets originates in the analysis of
molecular motors \cite{FAP97} whose function can be explained by an interplay of a coherent
external force and a breaking of binary symmetries like parity and time reversal. Removing macroscopic features like dissipation and noise culminated in the notion of Hamiltonian ratchets
\cite{FYZ00,Mat00,DF&01,SDK05} where the conservation of phase-space volume severely
restricts the generation of currents and time-reversal invariance (TRI) has to be broken by a
correspondingly asymmetric time dependence of the external force. As Hamiltonian systems,
they are readily quantized. Quantum ratchets reveal in particular relationships between
nonlinear transport and band structure \cite{SDK05}.

Restricting a ratchet to a finite compact space results in a pump \cite{Bro98,AG99}
a periodically forced scattering system that generates directed transport by an asymmetry in the
transmission and reflection coefficients, such that there is an overall bias for transport in one
direction \cite{DGS03,CDS12}. They can be implemented in particular in nanosystems such as
quantum dots \cite{KJ&91} or tunnel junctions \cite{PL&92}. Quantum effects in Hamiltonian pumps,
beyond the regime of adiabatic driving \cite{Bro98}, are appropriately described in the framework of
Floquet scattering theory \cite{CDS12,HDR01}.

\begin{figure}[h!]
\begin{center}
\includegraphics[width=13cm]{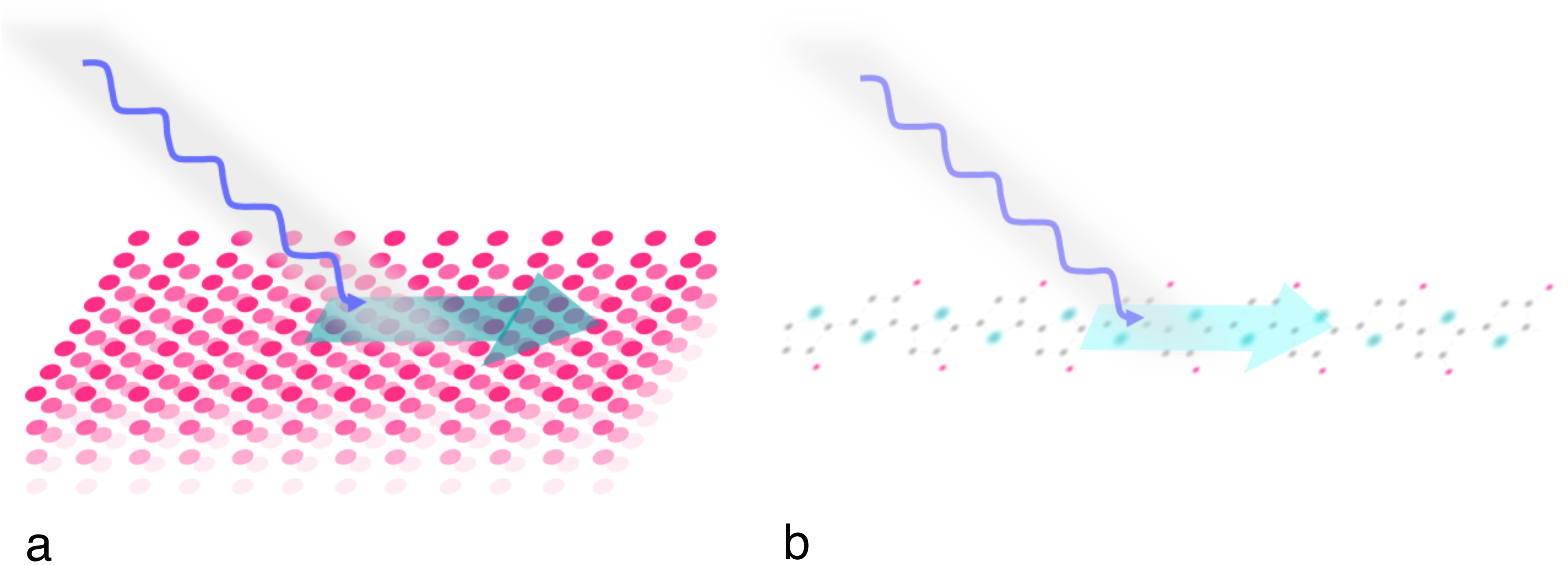}
\caption{Both in surfaces of crystalline metals or semiconductors (a) and in quasi-one-dimensional
molecules like polymers (b), a local coherent forcing, e.g., by a laser beam (blue wavy lines),
can induce directed currents (green arrows). They combine a spatially periodic static potential,
possibly asymmetric under reflection, as in ratchets, with a local periodic driving that breaks
time-reversal invariance, as in pumps.}
\label{figpolymetal}
\end{center}
\end{figure}

Attractive candidates for the study of nonlinear transport mechanisms are substances with
a periodic (crystalline) potential that are locally subjected to a coherent irradiation. Examples
are found in metal or semiconductor surfaces pumped by a collimated laser
(Fig.\ \ref{figpolymetal}a) \cite{HK&97,GR&07,GH08} as well as in polymers illuminated
locally by a cw laser (Fig.\ \ref{figpolymetal}b) \cite{Mur13}. These systems have their static
potential structure in common with ratchets but share the driving
restricted to a compact region in space, comprising, say, a single or a few unit cells
of the static potential, with pumps. We suggest the term ``punchets'' for this kind of
hybrid systems.

However, neither one of the theoretical approaches developed for pumps and ratchets are
in themselves sufficient to describe this system class. It is the purpose of this work
to propose an appropriate description of punchets in terms of driven scattering 
between asymptotic regions supporting periodic, as opposed to free, motion, to apply it to
a number of simple models, and to study in particular the currents generated in these systems.
Combining periodic potentials with a spatially confined driving, punchets offer a richer choice
of parameters and more options to control transport than the two system classes forming
their parents.

Systems combining features of ratchets and pumps allow for different schemes of
``division of labour'' between the two components, giving rise to a number of distinct types
of models. Closest to pumps are systems where it is exclusively the driving that breaks
temporal as well as spatial inversion symmetries. In this case, the ratchet character
comes in solely by the spatial periodicity of the static potential. However, it need
not by itself prefer any direction of transport, which then depends only on the more easily
controllable laser. By contrast, if the static potential does have an inherent directionality,
such as in a typical ratchet, the driving is only required to break time-reversal invariance (TRI),
as can be accomplished even with a beam perpendicular to the surface.
This means that the current, including its direction, can be controlled by manipulating solely
the spectral composition of the laser, without affecting the spatial configuration of the set-up.

The impact of a slanted laser beam, in turn, is more realistically modelled
by a running wave in the illuminated surface, with an envelope given by the
cross-section of the intensity profile in the plane of the surface. As a minimal
version modelling a propagating wave, two spatially separated drivings can be
considered, with a phase shift $\varphi \bmod 2\pi \neq 0,\pi$ between their respective periodic
time dependences. In this model class, a question of particular interest is whether
or not the direction of the induced current coincides with that of the wave propagation.

We shall begin in Section \ref{models} with the construction of simple
one-dimensional models for each of the three categories outlined above.
The dynamics generated by these models will be studied in Section \ref{scat},
focusing on evidence of mixed scattering, with integrable and irregular motion
coexisting in their phase space. As a more global property, directed transport
generated in these models is the subject of Section \ref{trans}, in particular
features such as current reversals that demonstrate its nonlinear character. We shall
summarize our results and indicate options of follow-up research, in particular quantum
punchets, in Section \ref{conc}.

\section{Three types of punchets}\label{models}

We here adopt the established basic construction elements of models for ratchets and
pumps, that is, a spatially periodic potential depending on the coordinate where
transport occurs, subject to a periodic driving force restricted to the support of
an envelope function that vanishes outside a compact interval. The overall form
of the Hamiltonian is therefore
\begin{equation}\label{genham}
H(p,x) = {p^2 \over 2m} + V_{\rm sta}(x) + V_{\rm env}(x) f_{\rm dri}(x,t),
\end{equation}
with a periodicity $V_{\rm sta}(x+L) = V_{\rm sta}(x)$ with period $L$ in space 
and $f_{\rm dri}(x+T,t) = f_{\rm dri}(x,t)$ with period $T$ in time. In all that follows,
$L = 2$, $T = 2$, and we choose the function
\begin{equation}\label{drienv}
V_{\rm env}(x) = \left\{\begin{array}{ll}
V_{{\rm env}\,0} \left(1+\beta{x \over b}\right)
\exp\left[{-1 \over (x^2 - a^2)^2}\right] & |x| < a,\\
0 & |x| \geq a .
\end{array}\right.
\end{equation}
for the envelope. It vanishes outside the interval $[-a,a]$ but is infinitely often
differentiable. The parameter $\beta$, $-b < \beta < b$, controls the degree
of asymmetry of the envelope, for $\beta = 0$ it is symmetric. In the following, $a = 2$ and $b = 1$.
Specifying further details for $V_{\rm sta}(x)$ and $f_{\rm dri}(x,t)$ will define the three
model categories announced above.

\begin{figure}[h!]
\begin{center}
\includegraphics[width=15cm]{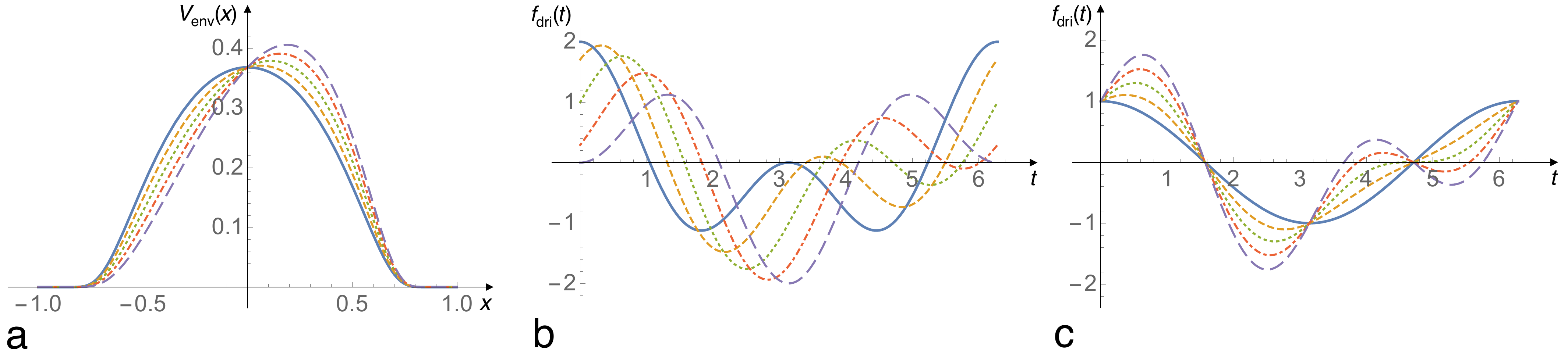}
\caption{Symmetry breaking in the $x$ and $t$ dependence of the driving function:
(a) spatial envelope of the driving, Eq.~(\protect\ref{drienv}), with $\beta = 0$
(full blue line), $0.25$ (orange dashed), $0.5$ (green dotted), $0.75$ (red dot-dashed),
$1.0$ (violet long-dashed), (b) time dependence, Eq.~(\protect\ref{dripot}),
at fixed $\alpha = 1$, for $\phi = 0$ (full blue line), $\pi/4$ (orange dashed), $\pi/2$ (green dotted),
$3\pi/4$ (red dot-dashed), $\pi$ (violet long-dashed), and at fixed $\phi = \pi/2$ (c),
for $\alpha = 0$ (full blue line), $0.25$ (orange dashed), $0.5$ (green dotted),
$0.75$ (red dot-dashed), $1.0$ (violet long-dashed).}
\label{figenvosc}
\end{center}
\end{figure}

\subsection{Models type 1: Only the driving breaks space and time symmetries}\label{typeone}

If all the asymmetries required to enable directed transport are left to the driving
force, we have to choose its position and time dependences so as to break both spatial
reflection and time reversal invariance. While the envelope (\ref{drienv}) for $\beta \neq 0$
already violates parity, we choose the time dependence as
\begin{equation}\label{dripot}
f_{\rm dri}(x,t) =  f_{\rm dri}(\omega t) = \cos(\omega t) + \alpha\cos(2\omega t - \phi),
\end{equation}
where $\omega = 2\pi/T$. Symmetry breaking is controlled by $\alpha$ and $\phi$,
such that for $\alpha = 0$ and for $\phi \bmod 2\pi = 0,\pi$, TRI is not broken. 
A laser driving of this type could be achieved producing a complementary beam,
to represent the $2\omega$-term in Eq.~(\ref{dripot}), by second-harmonic generation
and superposing the two beams with a controlled phase shift. Examples of
$V_{\rm env}(x)$ and $f_{\rm dri}(t)$ for different degrees of
symmetry breaking are depicted in Fig.~\ref{figenvosc}. Since with
Eqs.~(\ref{drienv},\ref{dripot}), space and time dependences factorize,
a propagating wave cannot be modelled in this way. 

It suggests itself to use the same form for the position dependence of the static
potential as in the time-dependent $f_{\rm dri}(t)$,
\begin{equation}\label{stapot}
V_{\rm sta}(x) = V_{{\rm sta}\,0} [\cos(qx) + \gamma\cos(2qx - \psi)].
\end{equation}
where $q = 2\pi/L$. As before, symmetry is recovered for $\gamma = 0$ and for 
$\psi \bmod 2\pi = 0,\pi$.

\subsection{Models type 2: sawtooth potential with time-asymmetric driving}\label{typetwo}

We here delegate the breaking of spatial reflection symmetry to the static crystal potential,
choosing the parameter in Eq.~(\ref{stapot}) accordingly, that is, $\gamma > 0$ and
$0 < \psi < \pi$. At the same time, the time dependence must break TRI, which requires
$\alpha > 0$ and $0 < \phi < \pi$, while the spatial envelope of the driving can be left symmetric,
$\beta = 0$ in Eq.~(\ref{drienv}). Additional control features are achieved, however,
if we also allow for $\beta \neq 0$, see Subsection \ref{currev}.

\subsection{Models type 3: localized propagating waves}\label{typethree}

A more sophisticated type of model is achieved if we allow for propagating waves under
the envelope (\ref{drienv}). This requires to give up the factorization between space- and
time-dependent modulation, introducing an $x$-dependence in Eq.~(\ref{dripot}).
At the same time, running waves already possess an inherent direction so that any further
symmetry breaking is not necessary. A plausible choice for the driving is
\begin{equation}\label{driwave}
f_{\rm dri}(x,t) =  f_{\rm dri}[k(x - v_\parallel t)],
\end{equation}
with a profile $f_{\rm dri}(\xi) = \cos(\xi) + \alpha\cos(2\xi - \phi)$ that can be asymmetric, as in
Eq.~(\ref{dripot}). The propagation velocity $v_\parallel$, the projection of the velocity
$\mathbf{v}_0$ of the incident beam to the plane parallel to the transport direction, is
\begin{equation}\label{vpar}
v_\parallel = {|\mathbf{v}_0| \over \sin(\theta)},
\end{equation}
if the beam is inclined by an angle $\theta \neq 0$ with respect to the surface normal.
Therefore, waves propagate in the negative $x$-direction if $\theta$ is positive and
{\em vice versa}. Evidently, for $\theta = 0$, a pure time dependence
$f_{\rm dri}(x,t) = f_{\rm dri}(\omega t)$ is retained, and the model reduces to the scheme of type 2.

A propagating wave already breaks space and time symmetries simultaneously. However, choosing an asymmetric shape for the static potential, such as in Eq.~(\ref{stapot}),
provides us with a another asymmetry parameter that can be controlled independently of
$\theta$. As we shall show in subsection \ref{currev}, this allows us to enforce current reversals
in a systematic manner and even transport against the direction of the running wave.

\begin{figure}[h!]
\begin{center}
\includegraphics[width=15cm]{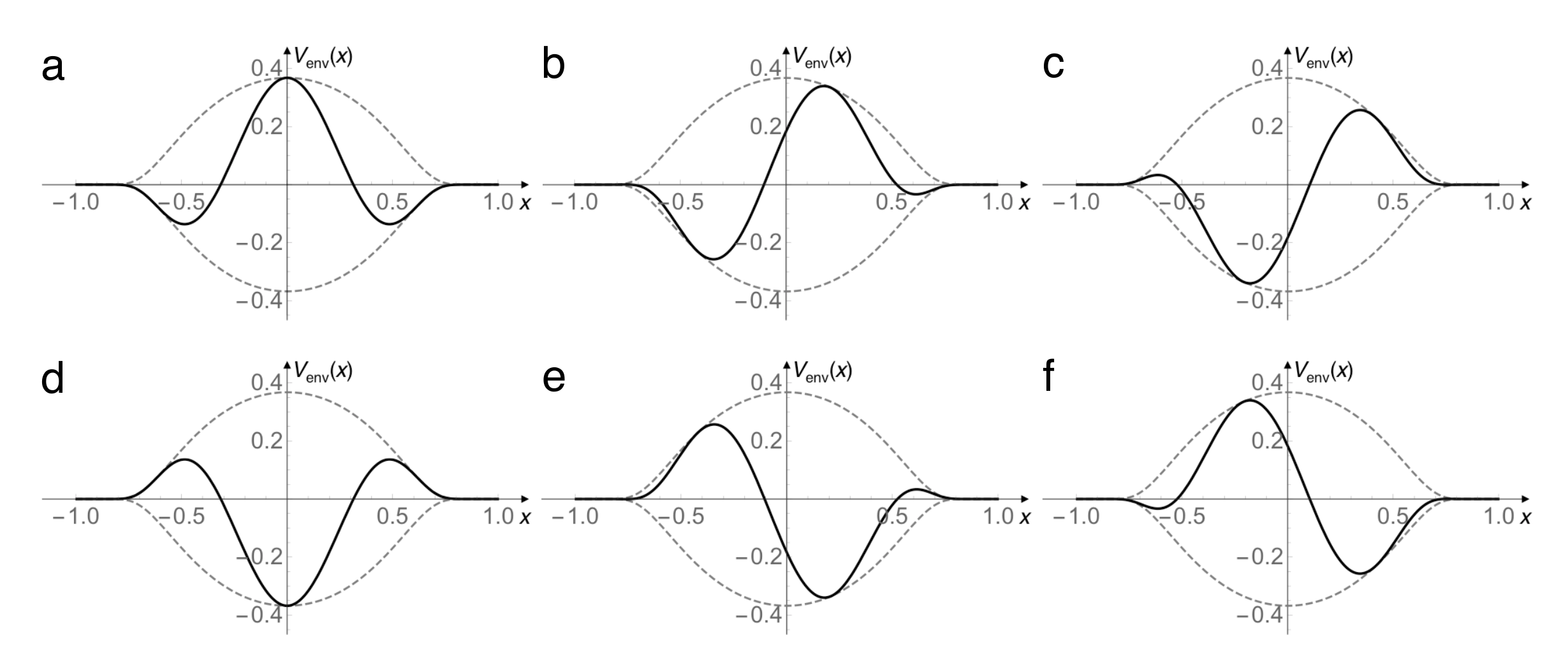}
\caption{Driving force in the form of a running wave (full lines) confined by a static envelope
(dashed), Eqs.~(\protect\ref{drienv}) and (\protect\ref{driwave}), at phases $kv_\parallel t = 0$ (a),
$\pi/3$ (b), $2\pi/3$ (c), $\pi$ (d), $4\pi/3$ (e), $5\pi/3$ (f).}
\label{figrunwave}
\end{center}
\end{figure}

As a rudimentary model of a running wave, the propagation range can be reduced to two
regions driven in synchrony but with a phase shift between them. Defining two envelopes,
\begin{equation}\label{twoenv}
V_{{\rm env}\,i}(x) = V_{\rm env}(x-x_i),\quad i = 1,2,
\end{equation}
centered at distinct points $x_1 \neq x_2$, harmonic driving functions
\begin{equation}\label{twowave}
f_{{\rm dri}\,i}(t) =  \cos(\omega t - \phi_i),\quad i = 1,2, \quad \phi_1 \neq \phi_2,
\end{equation}
are sufficient, provided the phase difference $\phi_2 - \phi_1$ is not an integer
multiple of $\pi/2$. 

\section{Irregular scattering}\label{scat}

In order to apply classical scattering theory to punchets, some generalization is
necessary. Including periodically driven scatterers already requires
to introduce an additional scattering parameter, the relative phase between the incoming
trajectory and the driving force \cite{DGS03}. It is defined as the phase
angle of the driving at the time when the extrapolated incoming asymptote reaches a
reference point within the scattering region, say, the origin $x = 0$. Similarly, if
motion in the asymptotic regions is not free but subject to a spatially periodic potential force, 
the phase of the incoming trajectory with respect to the phase of the static potential
is a relevant scattering parameter. These phases are hardly controllable in the
laboratory. In all that follows, we shall therefore consider averages over both of them.
That is, ensembles of initial conditions for the calculation of transport quantities,
in particular of currents comprise homogeneous distributions of initial times over one period
$T$ of the driving and of initial positions over one spatial period $L$ of the static potential.

In the presence of a periodic potential in the asymptotic regions, the definition of the incoming momentum has to be adapted as well.
Under this condition, phase space comprises two components, closed periodic orbits
trapped in the minima of the potential in each unit cell and traveling trajectories which
jump from cell to cell. They must have an energy above the absolute maximum of the potential,
\begin{equation}\label{asympene}
E_{\pm\infty} > \sup_{x_{\pm\infty} \leq x \leq x_{\pm\infty}+L} \bigl(V_{\rm sta}(x)\bigr),
\end{equation}
where $x_{-\infty}$, $x_{+\infty}$ are arbitrary reference points in the incoming or outgoing
asymptotic regions, resp. The time $T(E)$ a trajectory needs to pass across a single unit cell
is then a well-defined quantity, and we can introduce the asymptotic mean momentum as
\begin{equation}\label{asympmom}
\bar{p}_{\pm\infty}(E_{\pm\infty}) = m {L \over T(E_{\pm\infty})}.
\end{equation}
Based on these parameters, quantities and functions characterizing the scattering process
can be determined, in order to classify it between the extremes of integrable and chaotic
scattering \cite{Smi92}.

\begin{figure}[h!]
\begin{center}
\includegraphics[width=15cm]{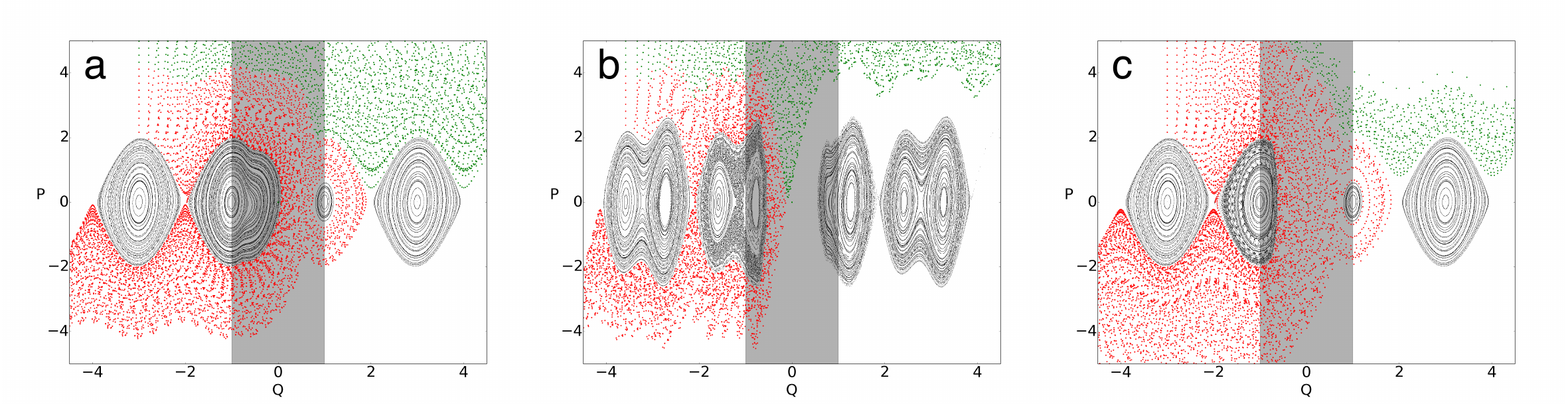}
\caption{Stroboscopic plots of the scattering region (gray) for the punchets
defined by Eqs.~(\protect\ref{genham}-\protect\ref{vpar}), showing trajectories
at integer multiples of the driving period, $t_n = nT$,
$n \in \mathbb{Z}$, $T = 2\pi/\omega$, for models of type 1 (panel a), type 2 (b),
and type 3 (c). Included are trapped trajectories (black) as well as traveling ones,
transmitted from left to right (green) and reflected from left to left (red). Parameters are 
$V_{{\rm sta}\,0} = 1, V_{{\rm env}\,0} = 20$, 
and $\alpha = \beta = 1$, $\gamma = \psi = 0$, $\phi = \pi/3$ (a),
$\alpha = \gamma = 1$, $\beta = 0$, $\phi = \psi = \pi/3$ (b),
$\alpha = 1$, $\beta = \gamma = 0$, $\phi = \psi = \pi/3$, $\theta = \pi/4$,
$k = 2.48337$, $v_0 = 2$ (c).}
\label{figstrobo}
\end{center}
\end{figure}

Of the criteria for irregular scattering proposed in Ref.~\cite{Smi92},
\begin{itemize}
\item[1] Poincar\'e surfaces of section show chaotic phase-space structures,
\item[2] deflection functions contain self-similar sections and in particular singularities
at a set of points with fractal dimension,
\end{itemize}
we here only verify the first one. (A third criterion often cited in the literature, exponential decay
of  the dwell-time distributions at least over a certain range of time scales, applies
in most cases but not unconditionally.) In Fig.~\ref{figstrobo}, we show stroboscopic plots
of the scattering region with adjacent unit cells, for models 1
(Fig.~\ref{figstrobo}a), 2 (Fig.~\ref{figstrobo}b), and 3 (Fig.~\ref{figstrobo}c).
They are analogous to Poincar\'e surfaces of section, as they replace intersections with a
reference surface in phase space by passages through reference points in the time domain,
typically synchronous with the driving, $t_n = nT$, $n \in \mathbb{Z}$. Coloured dots
indicate whether the trajectory is trapped in one of the minima of the potential (black),
or transmitted through the scattering region (green), or reflected (red).

\section{Directed transport}\label{trans}

\subsection{From asymmetric scattering to directed currents}\label{dircur}

\begin{figure}[h!]
\begin{flushright}
\includegraphics[width=13cm]{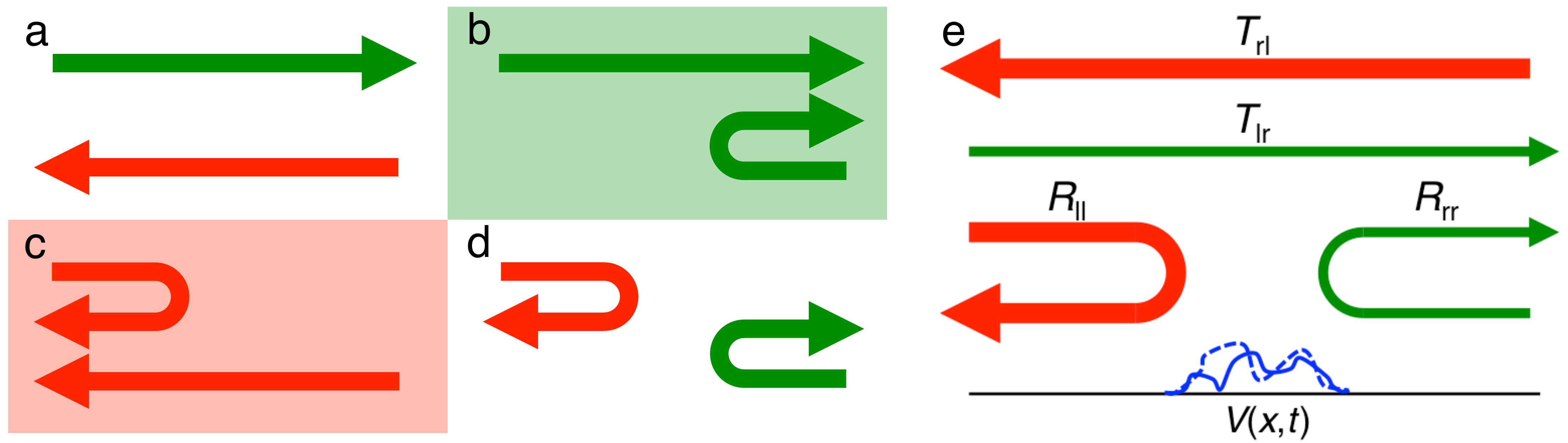}
\caption{From asymmetric scattering to directed transport. For an individual scattering event
(panels a to d) in an asymmetrically time-dependent scattering potential, all combinations
are possible, of transmission or reflection from the left side (${\rm l} \to {\rm r}$ or ${\rm l} \to {\rm l}$)
with transmission ${\rm r} \to {\rm l}$ or reflection ${\rm r} \to {\rm r}$ from the right side. A transport
imbalance occurs only if transmission from l combines with reflection from r (green arrows, b),
resulting in net transport to the right, or if reflection from l combines with transmission from r
(red arrows, d), resulting in net transport to the left. Averaged over ensembles of scattering events,
this may lead to asymmetric scattering coefficients (e), e.g.,
$T_{\rm rl} + R_{\rm ll} > T_{\rm lr} + R_{\rm rr}$, such that a non-zero total current
${\rm r} \to {\rm l}$ occurs.}
\label{figasymscat}
\end{flushright}
\end{figure}

In analyzing directed transport in terms of asymmetric scattering, we can adopt the reasoning
developed in the context of pumps. We define normalized currents $I_{\rm r}$ towards the right
and $I_{\rm l}$ towards the left as
\begin{equation}\label{currscat}
I_{\rm r} = T_{\rm lr} + R_{\rm rr}, \quad I_{\rm l} = T_{\rm rl} + R_{\rm ll},
\end{equation}
resp., denoting by $T_{\rm lr}$ the probability of transmission from left to right, by $ R_{\rm rr}$
the reflection probability from right back to right, etc. Normalizing the total {\em incoming} probability requires that $T_{\rm lr} + R_{\rm ll} = T_{\rm rl} + R_{\rm rr}$ or equivalently,
$T_{\rm lr} - T_{\rm rl} = R_{\rm rr} - R_{\rm ll}$. However, the {\em outgoing} probabilities
to the left and to the right need not be normalized individually. In the case of a time-dependent
driving, it is well possible that scattering is asymmetric under spatial reflection, and that accordingly
\begin{equation}\label{asymscat}
T_{\rm lr} + R_{\rm rr} \neq T_{\rm rl} + R_{\rm ll}.
\end{equation}
In that case, the total current from left to right,
\begin{equation}\label{totcurr}
I_{\rm tot} = I_{\rm r} -  I_{\rm l} = T_{\rm lr}  - T_{\rm rl}+ R_{\rm rr} - R_{\rm ll},
\end{equation}
need not vanish, cf.\ Fig.~\ref{figasymscat}.

\begin{figure}[h!]
\begin{center}
\includegraphics[width=15cm]{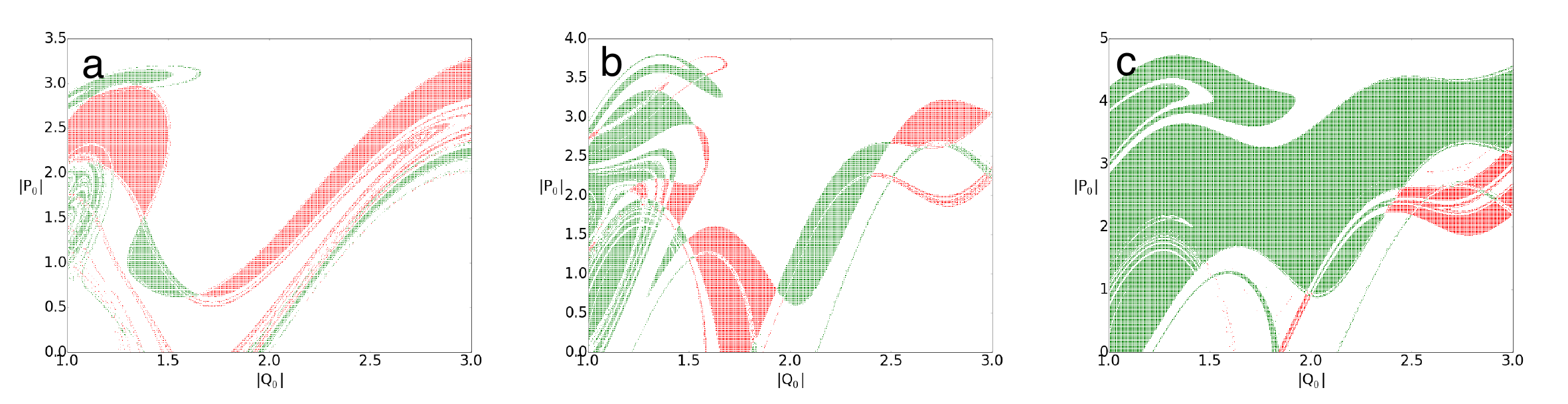}
\caption{Contribution of individual scattering events to transport. The colour of each
pixel indicates the outcome of a scattering process with initial momentum $p_0$
and position $q_0$, as indicated in Fig.~\protect\ref{figasymscat}a-d: transport
to the right, red (Fig.~\protect\ref{figasymscat}b), to the left, green
(Fig.~\protect\ref{figasymscat}c), or no transport, white (Figs.~\protect\ref{figasymscat}a,d).
Panels a, b, and c correspond to models 1, 2, and 3, resp. (see text). Parameters are 
$V_{{\rm sta}\,0} = 1, V_{{\rm env}\,0} = 20$, $\omega = \pi$,
and $\alpha = \beta = 1$, $\gamma = \psi = 0$, $\phi = \pi/2$ (a),
$\alpha = \gamma = 1$, $\beta = 0$, $\phi = \pi/2$, $\psi = \pi/3$ (b),
$\alpha = \beta = 0$, $\gamma = 1$, $\phi = 0$, $\psi = \pi/3$, $\theta = \pi/4$, $k = 2.48337$,
$v_0 = 2$ (c).}
\label{figcowplots}
\end{center}
\end{figure}

We can resolve transport mechanisms further, considering the contribution
of single scattering events (Fig.~\ref{figasymscat}a-d). On this level, asymmetry
becomes manifest as distinct outcomes, depending on the direction of the incoming trajectory,
with all other parameters kept the same. A net current to the right results if there
is transmission from left to right but reflection from right to right (Fig.~\ref{figasymscat}b),
to the left if transmission from right to left coincides with reflection from left to left 
(Fig.~\ref{figasymscat}c), but there is no net transport in the ``diagonal'' cases of transmission
(Fig.~\ref{figasymscat}a) or reflection (Fig.~\ref{figasymscat}d) from either side, always under
identical conditions except for the sign of the incoming momentum. We show data on asymmetry
in individual scattering events in Fig.~\ref{figcowplots}, using the colour code for each pixel as
indicated in Fig.~\ref{figasymscat}a-d. Obviously, if any of the binary symmetries impeding
transport are retained in the static potential or driving, plots would be void.
As the most conspicuous feature, we observe fractal structures in the transport properties,
evidence of the irregular nature of the underlying scattering processes. As
there is no reason for the area covered by one colour to exactly balance that covered by the other,
averaging transport over these contributions will generally result in asymmetric
scattering coefficients, hence in non-zero currents in Eq.~(\ref{totcurr}).

\begin{figure}[h!]
\begin{center}
\includegraphics[width=15cm]{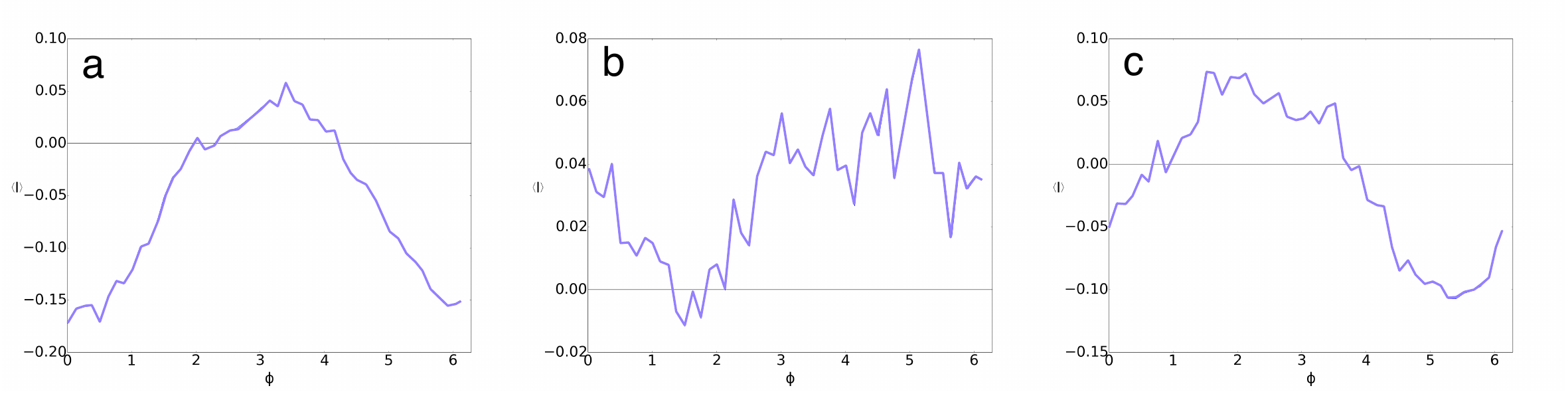}
\caption{Currents as functions of asymmetry parameters for models type 1 (a), type 2 (b),
and 3 (c). Normalized currents as defined in Eq.~(\protect\ref{totcurr}) are plotted as functions
of the asymmetry angle $\phi$ of the driving, cf.\ Eq.~(\protect\ref{dripot}) for symmetric
static potential ($\psi = 0$ in Eq.~(\protect\ref{stapot}), panel a), for asymmetric static potential
($\psi = \pi/3$) but spatially symmetric driving (b), and of the angle of incidence $\theta$,
cf.\ Eqs.~(\protect\ref{driwave},\protect\ref{vpar}), of a running wave (c). Other parameters are as
in Fig.~\protect\ref{figcowplots}, but with $\beta = -1$ for panel a.}
\label{figcurrentsgral}
\end{center}
\end{figure}

Computing currents as averages over phase space and possibly additional parameters
is a well-defined operation since the relevant momentum range is bounded. For
sufficiently high kinetic energy, the scattering potential can be neglected, so that scattering
reduces to transmission in either direction (Fig.~\ref{figasymscat}a) and no transport
occurs. If the bounds are $p_{{\rm max}\,-}$, $p_{{\rm max}\,+}$, resp., averages
can be restricted to the ranges $p_{{\rm max}\,-} < p < p_{{\rm max}\,+}$ and
$x_0 < x < x_0 + \bar{p}_{\pm\infty} T/m$, cf.\ Eqs.~(\ref{asympene},\ref{asympmom}).
Examples of how the generated current depends on the symmetry parameters
relevant for each of the three models are shown in Fig.~\ref{figcurrentsgral}.

\subsection{Symmetries and current reversals}\label{currev}

The symmetries inherent in our models, both in space and time, become manifest even on the
level of the generated currents and can be exploited as controlled ways to achieve, in particular,
current reversals. The parameters controlling the asymmetry of the static potential and the driving
play quite different roles. The prefactors $\alpha$ and $\gamma$ in Eqs.~(\ref{dripot}) and
(\ref{stapot}), resp., merely measure the magnitude of the symmetry-breaking terms and can
be kept positive. By contrast, the parameters $\beta$ in Eq.~(\ref{drienv}), $\psi$ in
Eq.~(\ref{stapot}), and $\theta$ in Eq.~(\ref{vpar}), controlling parity, and $\phi$ in
Eq.~(\ref{dripot}), controlling TRI, can take either sign and are directly related to the direction of
transport.

\begin{figure}[h!]
\begin{center}
\includegraphics[width=15cm]{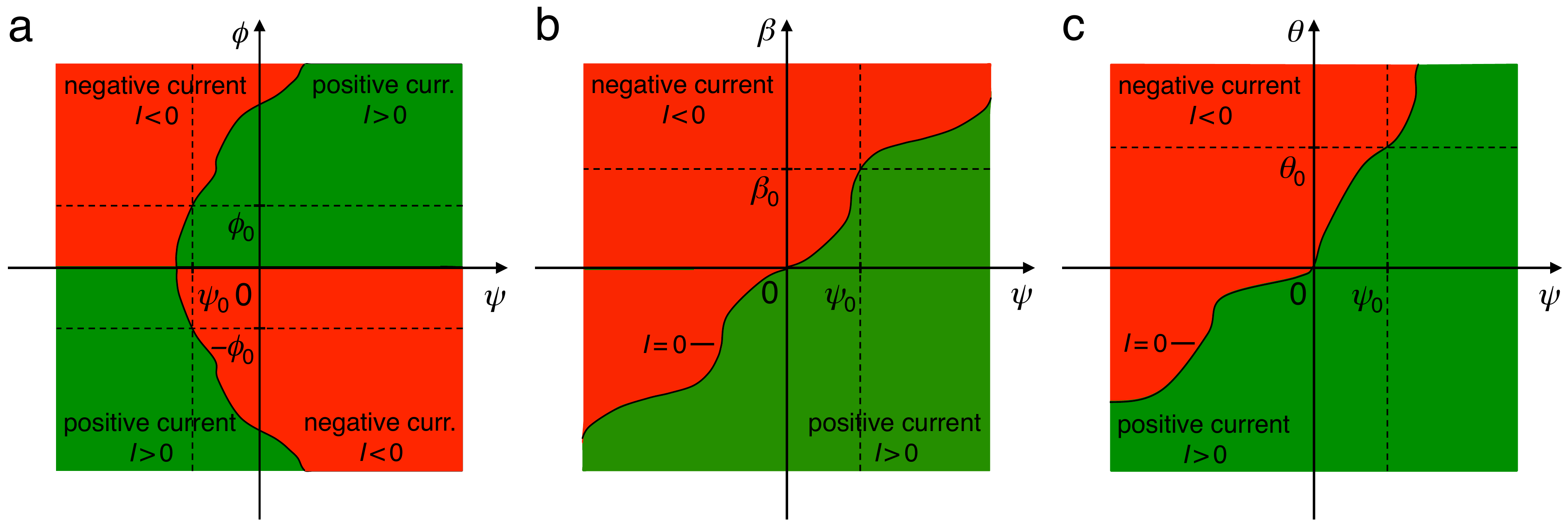}
\caption{Schematic transport phase diagrams for models of type 2 (a,b) and type 3 (c),
showing two-dimensional cross-sections of the three-dimensional parameter spaces
$(\beta,\psi,\phi)$ (a,b) and $(\theta,\psi,\phi)$ (c), at some fixed values of $\beta \neq 0$ (a),
of $\phi \neq 0$ (b), or of $\beta$ and $\phi$ (c).
Regions of  positive current (green) or negative current (red) are
indicated with the separating lines where $I = 0$ (black curves). For models of type 2 (a),
Eqs.~(\protect\ref{drienv},\protect\ref{dripot},\protect\ref{stapot}), at a fixed value $\beta > 0$,
directed transport requires $\phi \neq 0$ to break TRI. For each value $\phi_0 \neq 0$, the spatial
asymmetry controlled by $\psi$ implies that there will be a current reversal at some point
$\psi_0(\phi_0)$ and another, inverse reversal at the same $\psi_0$ but at $\phi = -\phi_0$.
Conversely, for a fixed value of $\phi \bmod \pi \neq 0$ so that TRI is broken (b), a
spatial asymmetry of the driving force implied by choosing $\beta_0 \neq 0$ can be
compensated by a simultaneous asymmetry of the static potential, leading to a current
reversal at some value $\psi_0 \neq 0$. The relation $I(-\beta,-\psi) = - I(\beta,\psi)$ is
reflected in the invariance of the borderline under this operation.
In models of type 3 (panel c), Eqs.~(\protect\ref{stapot},\protect\ref{driwave},\protect\ref{vpar}),
again at a fixed value $\beta > 0$, for sufficiently strong driving the current will always be in
the direction of the waves determined by the inclination $\theta$. If the driving competes
with a static ratchet potential with asymmetry $\psi_0$, a non-zero inclination
$\theta_0(\psi_0)$ is required to achieve a current reversal.}
\label{figseparatrix}
\end{center}
\end{figure}

In models of type 2, $\beta$ and $\psi$ provide two independent controls of the spatial asymmetry.
Inverting the sign of either one is equivalent to a reflection $x \to -x$. Under parity
$P:\;p \to -p,\;x \to -x,\;t \to t$, the direction of transport is reflected with $p$,
resulting in an inversion of the current,
\begin{equation}\label{currpar}
I(-\beta,-\psi,\phi) = -I(\beta,\psi,\phi),
\end{equation}
if $\phi$ is kept fixed or averaged over (Fig.~\ref{figseparatrix}b). ``Turning the crank the
other way round'', $\phi \to -\phi$, in turn, amounts to a time reversal $T:\;p \to -p,\;x \to x,\;t \to -t$
of the driving and likewise inverts the direction of transport,
\begin{equation}\label{currtri}
I(\beta,\psi,-\phi) = -I(\beta,\psi,\phi),
\end{equation}
not varying $\beta$ or $\psi$. As a consequence, there can be no current for $\phi = 0$
(Fig.~\ref{figseparatrix}a). The three-dimensional parameter space $(\beta,\psi,\phi)$ is therefore
traversed by two surfaces separating it into four quadrants of alternating sign of the current. One
of them coincides with the plane $\phi = 0$ and intersects the other along the $\phi$-axis
$\beta = \psi = 0$.

\begin{figure}[h!]
\begin{center}
\includegraphics[width=17cm]{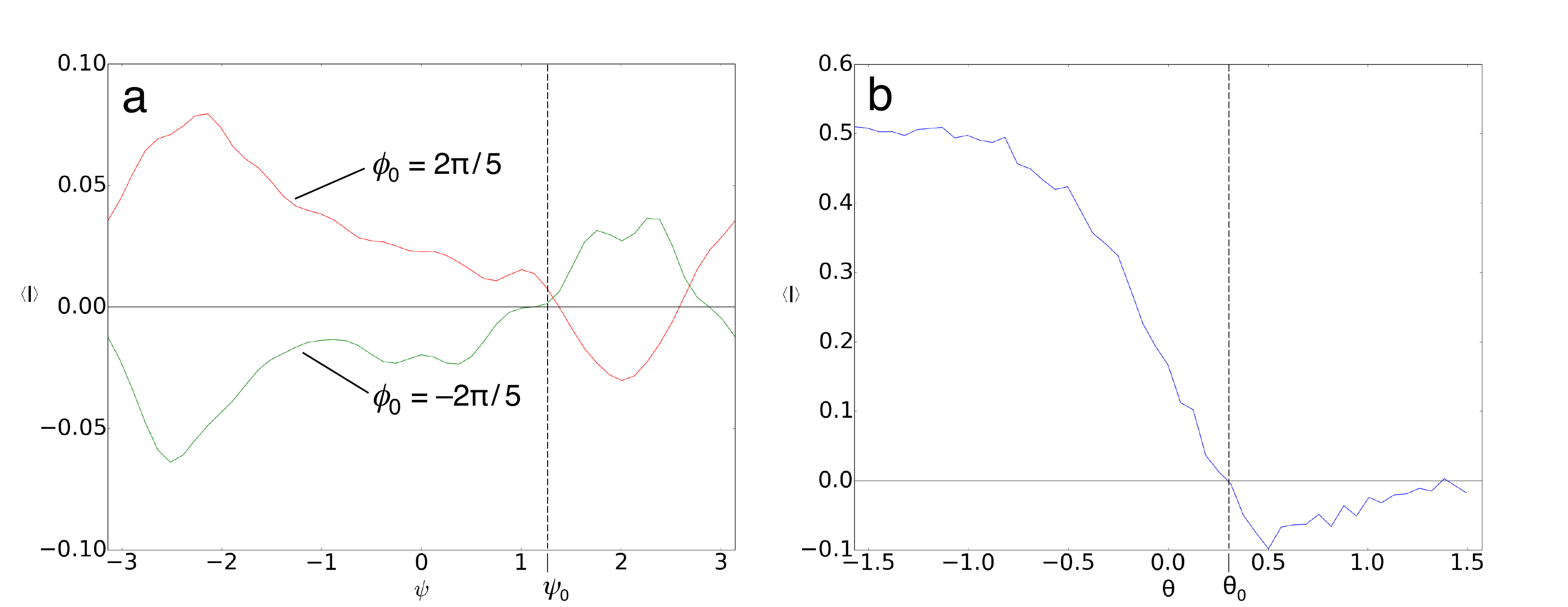}
\caption{Examples of current reversals, achieved systematically as outlined in the text and illustrated
in Fig.~\protect\ref{figseparatrix}. For model 2 (panel a), spatial asymmetry is controlled by
parameters $\psi$ and $\beta$ while $\phi$ controls the asymmetry under time reversal. Varying
$\psi$ at fixed $\beta_0 = 4$ and $\phi_0 = 2\pi/5$ (green), a current reversal is found at
$\psi_0 \approx 0.41\pi$ (dashed vertical line), as indicated in Fig.~\protect\ref{figseparatrix}a.
For $\phi_0 = -2\pi/5$ (red), the current globally changes sign, confirming Eq.~(\protect\ref{currpar}),
including an inverse current reversal at roughly the same value of $\psi_0$. Other parameters are
$V_{{\rm sta}\,0} = 1, V_{{\rm env}\,0} = 20$, $\omega = \pi/2$, and $\alpha = \gamma = 1$, $k = 2.48337$, $v_0 = 2$.
For a model of type 3 with asymmetric static potential (b),
the relevant two-parameter space is $(\psi,\theta)$. Here, $\theta$ is varied at fixed $\psi_0 = \pi/3$.
The current reversal (cf.\ Fig.~\protect\ref{figseparatrix}c) at $\theta_0 \approx 0.3$ is marked by a
dashed vertical line. For inclinations $\theta < \theta_0$, the current runs against the direction of the
driving wave. Other parameters are $V_{{\rm sta}\,0} = 6, V_{{\rm env}\,0} = 1$, $\omega = \pi$,
and $\alpha = 0.5$, $\phi = \pi/6$, $\beta = 0$, $\gamma = 1$, $k = 2.48337$, $v_0 = 2$.}
\label{figcurinv}
\end{center}
\end{figure}

For example, if transport towards the right is found for some fiducial values $\phi_0 \neq 0$ and $\beta \neq 0$ but at $\psi = 0$, varying $\psi$ sufficiently far in the direction
of decreasing magnitude of the current will typically lead to a point $\psi_0$ of vanishing current,
at the intersection with the separating surface related to $(\beta,\psi) \to (-\beta,-\psi)$
current reversals.

In models of type 3, the driving depends simultaneously on position and time through
the phase $k(x - v_\parallel t)$, cf. Eq.~(\ref{driwave}). Time reversal and parity cannot be
associated to separate parameter operations as in Eqs.~(\ref{currpar},\ref{currtri}). Instead,
it is a single operation that effects an exact inversion of the current,
\begin{equation}\label{currcri}
I(-\beta,-\psi,-\theta,-\phi) = -I(\beta,\psi,\theta,\phi),
\end{equation}
As a consequence, there is only a single surface that divides the $(\beta,\psi,\theta;\phi)$-space
in two halves of opposite sign of the current (Fig.~\ref{figseparatrix}c). It passes through the origin 
$\beta = \psi = \theta = \phi = 0$ but otherwise does not generally coincide with any of the axes.

At the same time, for sufficiently large amplitude $V_{{\rm env}\,0}$
of the waves, transport in their direction of propagation can always be enforced. 
This provides us with a particularly robust possibility to switch the current direction.
For a driving with perpendicular incidence ($\theta = 0$), an asymmetric static ratchet potential
(Eq.~(\ref{stapot}) with $\gamma > 0$ and $\psi \bmod 2\pi = \psi_0 \neq 0,\pi$) and/or an asymmetric envelope (Eq.~(\ref{drienv}) with $\beta \neq 0$), combined
with a time dependence of the driving that breaks TRI (Eq.~(\ref{dripot}) with $\alpha > 0$ and
$\phi \bmod \pi/2 \neq 0$), will induce transport as in models of type 2, say in the positive direction (Fig.~\ref{figseparatrix}c). If the incident beam is then inclined towards the right
by varying $\theta$, such that the waves propagate to the left, against the ratchet current,
the driving by the traveling wave competes with the ratchet effect and eventually 
supersedes it.

For a driving that is not strong enough to compensate
the ratchet effect completely, however, the induced current will maintain the direction of pure ratchet
transport, that is, {\em against} the running wave. In Fig.~\ref{figcurinv}, we show an example
of a current reversal enforced as outlined here, at an inclination
$\theta_0 \approx 0.3$. For more oblique incidence, $0 < \theta < \theta_0$, we find transport
against the direction of the waves inducing it.

\section{Conclusion}\label{conc}

In this paper, we introduce a novel class of models for nonlinear directed transport,
conceived as hybrids between ratchets and pumps. They combine a periodic driving
with an asymmetric time dependence, confined to a compact region in space,
with a static ratchet potential that breaks spatial reflection symmetry in the asymptotic
regions not affected by the driving. Within this framework, we discuss three different
types of models of increasing complexity, ranging from symmetric static potentials
with a driving that breaks both space and time symmetries to a driving by running waves
confined by a localized envelope.

The motion in these systems is characterized by irregular scattering, induced by the
periodic driving, between asymptotic regions where the motion is not ballistic but
invariant under discrete translations in space. Adopting methods developed
for directed transport in periodically driven scattering systems, we evaluate currents
by averaging over the outcomes of individual scattering events. Other parameters
kept equal, they may depend on the direction of the incident momentum and are reflected
in an imbalance of transmission and reflection coefficients from either side.

The resulting currents show the characteristic features of nonlinear directed transport,
such as a strong and often irregular parameter dependence and in particular the
phenomenon of current reversals. In the case of a driving by localized running waves,
it is even possible that the induced current runs against the propagation of the waves,
a striking manifestation of the nonlinear nature of this process. It shows that
punchets, as a synthesis of ratchets and pumps, offer a qualitatively wider range
of possibilities to achieve and control transport than either one of the parent models.

In this study of classical punchets, the complex motion underlying the transport
phenomena was in the focus. As a particular complementary research direction, 
the study of quantum punchets will face the quantum mechanical consequences
of the invariances involved, binary symmetries as well as periodicities. It requires to apply Floquet
scattering theory, as appropriate framework for periodically driven scattering, to Bloch states and
band spectra describing the quantum mechanics in the asymptotic regions.
We expect that the simultaneous restriction of scattering processes, by the
conservation of energy {\em modulo} the photon energy of the driving laser
and by the requirement to scatter only from band to band, will lead to a particularly
rich parameter dependence of the induced currents.

\section*{Acknowledgments}

We enjoyed inspiring discussions with Julio Arce and Gustavo Murillo (Universidad del Valle,
Cali, Colombia) and with Doron Cohen (Ben Gurion University, Beer Sheva, Israel).

\end{document}